\documentclass[12pt,preprint]{aastex}

\slugcomment{to be published in ApJ}

\shorttitle{Globular Cluster Multimodality from Radial Variation of CSPs}

\shortauthors{Pipino, Puzia, \& Matteucci}

\begin{document}

\title{The Formation of Globular Cluster Systems in Massive Elliptical Galaxies: Globular Cluster Multimodality from Radial Variation of Stellar Populations}





\author{Antonio Pipino\altaffilmark{1,3}, Thomas H. Puzia\altaffilmark{2,4}, \& Francesca Matteucci\altaffilmark{3}} 

\altaffiltext{1}{Astrophysics, University of Oxford, Denys Wilkinson Building, Keble Road, Oxford OX1 3RH, U.K. email: {\it axp@astro.ox.ac.uk}.} 

\altaffiltext{2}{Herzberg Institute of Astrophysics, 5071 West Saanich Road, Victoria, BC V9E 2E7, Canada, email: {\it puziat@nrc.ca}.}

\altaffiltext{3}{Dipartimento di Astronomia, Universit\`a di Trieste, 
Via G.B. Tiepolo 11, 34100 Trieste, Italy, email: {\it matteucci@ts.astro.it}.}

\altaffiltext{4}{Space Telescope Science Institute, 3700 San Martin Drive,
    Baltimore, MD 21218, USA.}

\begin{abstract}

The most massive elliptical galaxies show a prominent multi-modality in
their globular cluster system color distributions. Understanding the
mechanisms which lead to multiple globular cluster sub-populations is
essential for a complete picture of massive galaxy formation. By assuming
that globular cluster formation traces the total star formation and taking
into account the radial variations in the composite stellar populations
predicted by the Pipino \& Matteucci (2004) multi-zone photo-chemical
evolution code, we compute the distribution of globular cluster properties
as a function of galactocentric radius.~We compare our results to the
spectroscopic measurements of globular clusters in nearby early-type
galaxies by Puzia et al. (2006) and show that the observed multi-modality
in globular cluster systems of massive ellipticals can be, at least
partly, ascribed to the radial variation in the mix of stellar
populations. Our model predicts the presence of a super-metal-rich
population of globular clusters in the most massive elliptical galaxies,
which is in very good agreement with the spectroscopic observations. The
size of this high-metallicity population scales with galaxy mass, in the
sense that more massive galaxies host larger such cluster populations. We
predict an increase of mean metallicity of the globular cluster systems
with host galaxy mass, and forecast that those clusters that were formed
within the initial galaxy halo closely follow an age-metallicity and
$\alpha$/Fe-metallicity relation, where older clusters exhibit lower
metallicities and higher $\alpha$/Fe ratios. Furthermore, we investigate
the impact of other non-linear mechanisms that shape the metallicity
distribution of globular cluster systems, in particular the role of
merger-induced globular cluster formation and a non-linear
color-metallicity transformation, and discuss their influence in the
context of our model. We find that a non-linear color-metallicity relation
may be partly responsible for a color multi-modality. On the other hand,
the formation of globular clusters from subsequently accreted gas, either
with primordial abundances or solar metallicity, delivers model
predictions which are at variance with the observations, {and suggests that
a significant fraction of metal-poor globular clusters is accreted from
satellite halos which formed their globular clusters independently.}
\end{abstract}


\section{Introduction}

\subsection{The Multimodality of Globular Cluster Systems}
One of the most significant developments in the study of extragalactic
globular cluster systems (GCSs) was the discovery of bimodality in their
color distributions (see \citealt{ashman98, harris01, west04} and
references therein).~Today, we generally refer to globular clusters (GCs)
belonging to the blue peak of the color distribution as metal-poor GCs and
to the red-peak members as the metal-rich sub-population.~{ It is
generally considered that the presence of multiple modes implies multiple
distinct GC formation epochs and/or mechanisms and ties those directly
into formation scenarios that have to describe the {\it parallel} assembly
histories of GCSs {\it and} the diffuse stellar populations in their host
galaxies.~In massive early-type galaxies the current GCS assembly
paradigms view the origin of the two color peaks from the perspective of
either episodic star-cluster formation bursts triggered by gas-rich galaxy
mergers \citep[e.g.][]{ashman92}, temporarily interrupted cluster
formation \citep[so-called {\it in-situ} formation, e.g.][]{forbes97,
harris98}, and star-cluster accretion as a result of the hierarchical
assembly of galaxies \citep[e.g.][]{cote98}.}

While the majority of GCSs in early-type galaxies show clearly bimodal
color distributions, the general picture is much more complex, ranging
from purely blue to purely red color distributions \citep[e.g.][]{gk99,
kundu01a, kundu01b, larsen01, peng06}. This complexity is exacerbated by
the fact that color bimodality is a function of galaxy mass and
morphology, as less massive and later-type galaxies tend to have
single-mode blue (i.e. metal-poor) GC populations \citep[e.g.][]{lotz04,
sharina05, peng06}. Furthermore, color bimodality is also a function of
galactocentric distance and is mainly due to the more extended spatial
distribution of the metal-poor sub-population relative to metal-rich
clusters \citep[e.g.][]{harris02, rhode04, dirsch03, dirsch05}.

\subsection{Numerical Models of Globular Cluster System Formation}

The aspect of GCS formation and assembly entered
recently the domain of numerical simulations of galaxy formation due to
the increasing spatial resolution of these computations.~For instance,
\cite{li04} model GC formation by identifying absorbing sink
particles in their smoothed particle hydrodynamics (SPH) high-resolution
simulation of isolated gaseous disks and their mergers.~They find a
bimodal globular-cluster metallicity distribution in their
merger remnant under the assumption of a particular age-metallicity
relation. A key finding of their merger simulation is a more
concentrated spatial distribution of metal-rich GCs with
respect to the metal-poor sub-population in good agreement with
observations. Since their models of isolated galaxies produce a smooth age
distribution (implying a smooth metallicity and color distribution),
\citeauthor{li04} conclude that mergers are required to produce a bimodal
metallicity (i.e.~color) distribution.

In a more detailed adaptive-grid cosmological simulation,
\cite{kravtsov05} followed the formation of a star-cluster system during
the early evolution of a Milky Way-size disk galaxy to redshift
$z\!=\!3$.~Their model could reproduce the extended spatial distribution
of metal-poor halo globular clusters as observed in M31 and the Milky Way.
However, because their simulation does not follow the later evolution at
$z<3$ it is unclear whether it would produce a metallicity bimodality and
any significant age-metallicity relation.

An alternative, more statistical approach to modeling GCS assembly is to
directly link the mode of GC formation to the star-formation rate in
semi-analytic models. { \cite{beasley02} were the first to explore this
path by assuming that metal-poor GCs} form in gaseous proto-galactic disks
while metal-rich GCs are created during gaseous merger events.~Their study
showed that the observed globular-cluster color bimodality can only be
reproduced by artificially stopping the formation of metal-poor GCs {
at redshifts $z\!\ga\!5$}. By construction, no spatial information on
metal-rich and/or metal-poor GCs is provided in these models.

\subsection{A Spatially Resolved Chemical Evolution Model for Spheroid Galaxies} 

Recently, \citet[][hereafter PM04]{pipino04} presented a spatially
resolved chemical evolution model for the formation of spheroids, which
successfully reproduces a large number of photo-chemical properties that
could be inferred from either the optical or from the X-ray spectra of the
light coming from ellipticals. The model includes an initial gas infall
and a subsequent galactic wind; it takes into account detailed
nucleosynthesis prescriptions of both type-II and Ia supernovae { as
well as low and intermediate-mass stars}. It has been extensively tested
against the main photo-chemical properties of nearby ellipticals,
including the observed increase of the $\alpha$-enhancement in their
stellar populations with galaxy mass \citep[e.g.][]{worthey92, weiss95}.
This is at variance with standard models based on the hierarchical merging
paradigm, which do not reproduce this trend \citep{thomas02}.

Since the PM04 model provides full radial information on the {
composite nature} of stellar populations that make up elliptical galaxies,
the observation of different GC sub-populations is, therefore, a new
sanity check for the validity of this model.~Moreover, we recall that PM04
and, more recently, \citeauthor{pipino06} (\citeyear{pipino06}, hereafter
PMC06) suggested that elliptical galaxies should form outside-in, namely
the outermost regions form faster as well as develop an earlier galactic
wind with respect to the central parts \citep[see also][]{martinelli98}.
{This mechanism implies that the stars in massive spheroids form a
Composite Stellar Population (CSP), whose chemical properties, in
particular their metallicity distribution, changes with galactocentric
distance.}

Starting with the assumption that GC sub-populations trace the components
of CSPs, {we will show how the observed multi-modality in GCSs can be
ascribed} to the radial variation in the underlying stellar populations.
In particular, the observed GCSs are a linear combination of GC
sub-populations inhabiting a given projected galactocentric radius.

The paper is organized as follows: in Section 2 we briefly describe the
adopted theoretical model; in Section 3 we compare the predictions with
observations and discuss the implications, while Section 4 presents the
final conclusions.


\section{The model}


\subsection{The Chemical Evolution Code}

The chemical evolution code for elliptical galaxies adopted here is
described in PM04, where we refer the reader for more
details. In this work, we present the results for a galaxy with $M_{\rm
lum}\sim\!10^{11}M_{\odot}$, taken from PM04's Model IIb. This model is
characterized by a \cite{salpeter55} IMF, \cite{thielemann96} yields for
massive stars, \cite{nomoto97} yields for Type-Ia SNe, and \cite{vdH97}
yields for low- and intermediate-mass stars.

An important feature of the PM04 model is its multi-zone nature, namely
the model galaxy is divided into several non-interacting spherical shells
of radius $r_i$, which facilitate a detailed study of the radial variation
of the photo-chemical properties of the GCS and its
host galaxy. In each zone \emph{i}, the equations for the chemical
evolution of 21 chemical elements are solved (see PM04, \citealt{matteucci01}).

The model assumes that the galaxy assembles by merging of gaseous lumps on
short timescales. The chemical composition of the lumps is assumed to be
primordial. In fact, our model assumes that the accretion of primordial
gas from the surroundings\footnote{Since we lack a cosmological framework
we cannot further specify the properties of the infalling primordial gas.}
is more efficient in more massive systems, given their higher cross
section per unit mass (see PM04). The model galaxy suffers a strong
starburst which injects a large amount of energy into the interstellar
medium, able to trigger a galactic wind, occurring at different times at
different radii, mainly due to the radial variation of the potential well,
which is shallower in the galactic outskirts. 
After the onset of wind activity the star formation is
assumed to stop and the galaxy evolves passively with continuous mass
loss. In order to correctly evaluate the amount of energy driving the
wind, a detailed treatment of stellar feedback is included in the code
(that takes into account the stellar lifetimes). In particular, the energy
restored to the interstellar medium by both Type-Ia and Type-II supernovae
has been calculated in a self-consistent manner according to the time of
explosion of each supernova and the characteristics of the ambient medium
(see PM04 for details). The potential well that keeps the gas bound to the
galaxy is assumed to be dominated by a diffuse and massive halo of Dark
Matter surrounding the galaxy.

In the following we adopt the standard star formation rate $\psi_*
(t,r_i) = \nu \cdot \rho_{\rm gas}(r_i,t)$ before the onset of the
galactic wind ($t_{\rm gw}$), where $\rho_{\rm gas}$ is the gas density,
$\nu$ the star-formation efficiency; 
otherwise we assume that $\psi_*
(t>t_{\rm gw},r_i) = 0$.~We recall here that the adopted
star-formation efficiency is $\nu =10 \rm \, Gyr^{-1}$, while the infall
timescale is $\tau\! =\!0.4 \rm \, Gyr$ in the galactic core and $\tau =0.01
\rm \, Gyr$ at one effective radius (of the diffuse light, $R_{\rm eff}$),
respectively. These values were chosen by PM04 in order to reproduce the
majority of the chemical and photometric properties of ellipticals such
as: the $\langle$[Mg/Fe]$\rangle\!-\!\sigma$ \citep[e.g.][]{faber92}, the
Color-Magnitude \citep[e.g.][]{bower92}, the Mass-Metallicity relation
\citep[e.g.][]{gallazzi05} as well as the observed gradients in
metallicity \citep[e.g.][]{carollo93}, $\langle$[Mg/Fe]$\rangle$
\citep[e.g.][]{mendez05}, and color \citep[e.g.][]{peletier90}.
\cite{pipino05} recently extended this model to explain also the
properties of hot X-ray emitting halos surrounding more massive spheroids.


\subsection{Globular Cluster Formation}

The formation rate of GCs, $\psi_{GC}$, in the \emph{i-}th
shell is assumed to be directly linked to its star formation rate $\psi_*
(t,r_i,Z_i)$ via a suitable function of time \emph{t}, radius $r_i$, and metallicity
\emph{Z}, which represents some scaling law between star formation rate $\psi_*$
and the star cluster formation $\psi_{\rm GC}$ and can be regarded as a GC
formation efficiency. A similar relation between the average star
formation rate per surface area and the star cluster formation was
recently found by \cite{larsen00} to hold in nearby spiral galaxies.
In addition, the efficiency of cluster formation in massive ellipticals
appears to be constant, where the mass ratio between the mass in star
clusters and the baryons locked in field stars+gas is $\epsilon_{\rm
GC}\!\approx0.25\%$ \citep{mclaughlin99}. Here we extend this surface
density relation to 3-D space. 

Moreover, PMC06 showed that at a given galactocentric radius model galaxies are made
of a CSP, namely a mixture of several
simple stellar populations (SSPs) each with a single age and chemical
composition. The CSP reflects the chemical enrichment history of the
entire system, weighted by the star formation rate. We define the stellar
metallicity distribution $\Upsilon_*$ as the distribution of stars
belonging to a given CSP as a function of [Z/H]. 

We can then write the globular-cluster metallicity distribution
$\Upsilon_{\rm GC}$ at a given radius $r_i$ and time $t$ as:
\begin{equation}
\Upsilon_{\rm GC}(t,r_i, Z)=f(t,r_i,Z)\cdot \Upsilon_* (t,r_i,Z) \; ,
\end{equation}
where $f$ includes all the information pertaining to the
connection between $\psi_{GC}$ and $\psi_*$.

It is not trivial, and beyond the scope of
the paper, to find an explicit definition for $f(t,r_i,Z_i)$, which
basically carries the information on the internal physics of gas clouds
where GCs are expected to form. In the following we will show that, for a
few and sensible choices of $f$, the observed multi-modality in the color
distribution of globular clusters may be driven by the radial
variations in the stellar population mix of ellipticals. We will first
adopt a constant function $f$ (see Sec.~\ref{bimod}) and then allow $f$ to
mildly vary with $Z$ (see Sec.~\ref{ref:metals}). No absolute values for
$f$ will be given, since our formalism deals with normalized
distributions.

In particular, the \emph{total} $\Upsilon_{\rm GC}$ summed over all radial
shells can be written as:
\begin{equation}
\Upsilon_{\rm GC, tot}(t,Z)=\sum_i f(t,r_i,Z)\cdot \Upsilon_* (t,r_i,Z)\; .
\label{gcmd1}
\end{equation}
Similar equations hold for other GC distributions as a
function of either [Mg/Fe] or [Fe/H].

At this stage it is useful to recall that $\Upsilon_* (t,r_i,Z)$ can be
represented in two following ways: $i)$ as the fraction of mass of a CSP
which is locked in stars at any given metallicity \citep{pagel75,
matteucci01}. In the following we refer to this stellar metallicity
distribution as $\Upsilon_{*,m}$ (MSMD: mass-weighted stellar metallicity
distribution); $ii)$ as a fraction of luminosity of the CSP in each
metallicity bin. This definition is closer to the measurement as it can be
directly compared to the luminosity-weighted mean $\Upsilon_{*,l}$ (LSMD:
luminosity-weighted stellar metallicity distribution) at a given radius
\citep[see][]{arimoto87, gibson96}. This classification is important since
PMC06 showed that the $\Upsilon_{*,m}$ and $\Upsilon_{*,l}$ might differ,
especially at large radii, even for old stellar populations. {The
advantage of GCSs, for which accurate ages are known, is that they 
directly probe the mass-weighted distributions.}

{At this point it is useful to recall that our the adopted chemical evolution model divides
the galaxy in several non-interacting shells. In each shell the time
at which the galactic wind occurs is self-consistently evaluated
from the \emph{local} condition. In particular, we follow Martinelli et al. (1998) 
suggestion that gradients can arise as a consequence of a more prolonged SF, and thus
stronger chemical enrichment, in the inner zones. In the galactic
core, in fact, the potential well is deeper and the supernovae (SNe)
driven wind develops later relative to the most external regions.
This particular formation scenario leaves a
characteristic imprint on the shape of both  $\Upsilon_{*,m}$ and $\Upsilon_{*,l}$
and here we give some general considerations.
In particular, we can explain the slow rise in the low metallicity tail of the distributions 
as the effect of the initially infalling gas, whereas the onset of the galactic
wind sets the maximum metallicity of the $\Upsilon_{*,m}$ and $\Upsilon_{*,l}$.
In general, the suggested outside-in formation
process reflects in a more asymmetric stellar metallicity distribution 
at larger radii, where the galactic wind occurs earlier (i.e. closer to
the peak of the star formation rate), with respect to the galactic
center. The \emph{qualitative} agreement between these model predictions 
and the observed stellar metallicity distributions 
derived at different radii by Harris \& Harris (2002, see their
Fig. 18) 
for the stars in the elliptical galaxy NGC 5128 is remarkable.
If confirmed from observations in other ellipticals, the expected
sharp truncation of $\Upsilon_{*,m}$ at large
radii might be the first direct evidence of a sudden and strong wind
which stopped the star formation earlier in the galactic outskirts (see PMC06
and Pipino, D'Ercole, \& Matteucci, in preparation).}


\section{Results and discussion}


\subsection{The Multi-Modality of Globular Cluster Systems in Elliptical Galaxies}

\label{bimod}

The general presence of multi-modal GCSs implies that their host
galaxies did not form in a single, isolated monolithic event, but
experienced spatially and/or temporally separated star-formation
bursts. In the recent past, both semi-analytic and hydrodynamic
simulations of galaxy formation attempted to follow the process of GC
formation \citep{beasley02, kravtsov05}, but neither
could produce a clearly bimodal MDF in their simulated GCSs.

\begin{figure}[!ht]

\epsscale{1.1}
\plotone{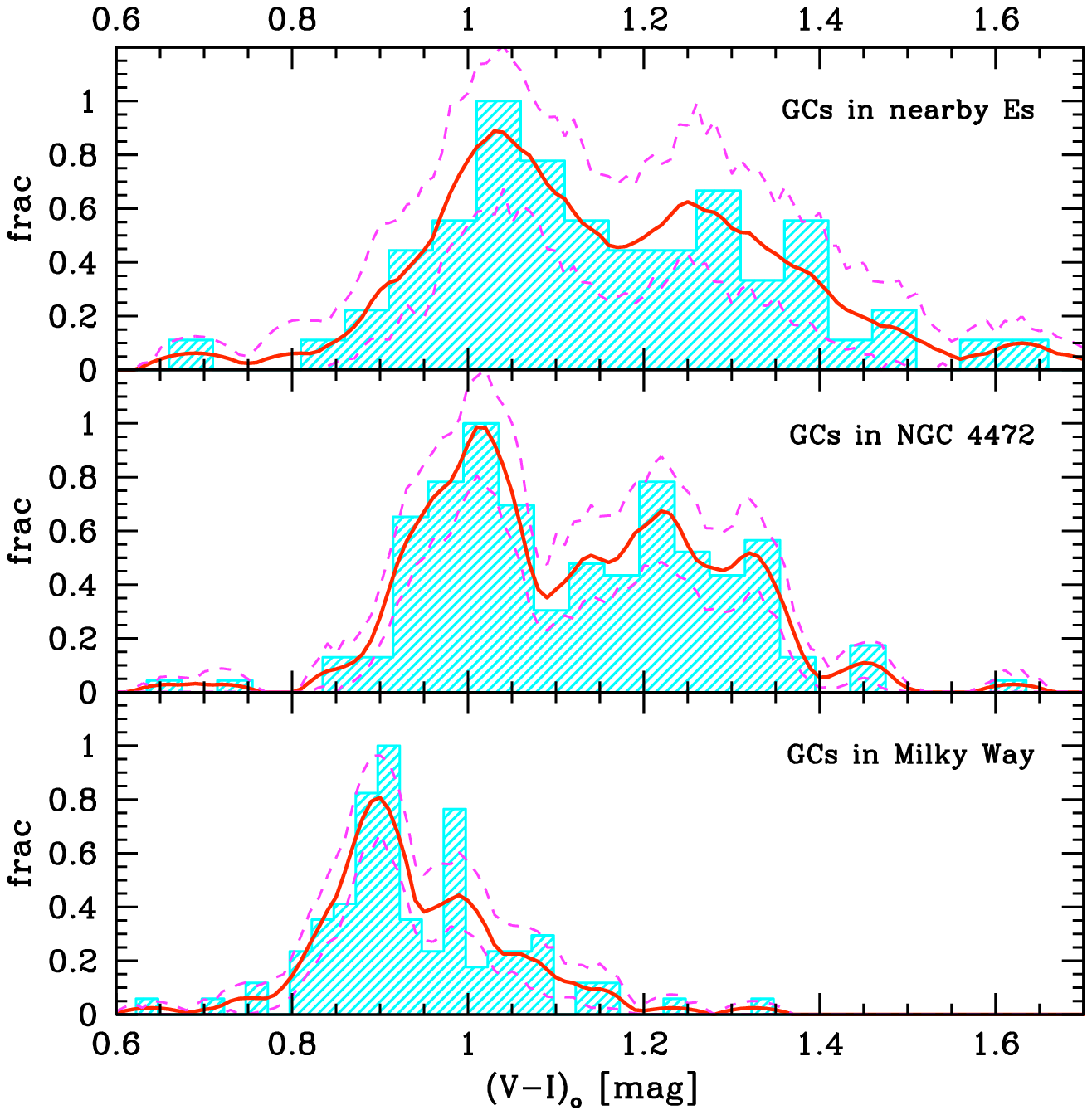}

\caption{Color distributions of GCs in nearby, massive elliptical galaxies
\citep[{\it top panel}]{puzia06}, in NGC~4472 \citep[][{\it middle
panel}]{puzia99}, and the Milky Way, taken from the February 2003 update
of the McMaster catalog \citep[][{\it bottom panel}]{harris96}. 
In order to allow a robust comparison between the P06 and NGC~4472 sample,
only GCs in NGC~4472 with with luminosities brighter than $V\simeq22.5$
mag are shown. The solid lines are Epanechnikov-kernel probability density 
estimates with their bootstrapped 90\% confidence limits.}

\label{colsamp}

\end{figure}

In this section we will show how to obtain a bimodal metallicity
distribution function for GCs $\Upsilon_{\rm GC, tot}$
starting from single-mode stellar metallicity distribution functions $\Upsilon_* (t,r_i,Z)$
(commonly known as G-dwarf-like diagrams) for the CSP inhabiting different
radii of a prototypical elliptical galaxy according to Equation~\ref{gcmd1}.

\begin{figure*}


\plottwo{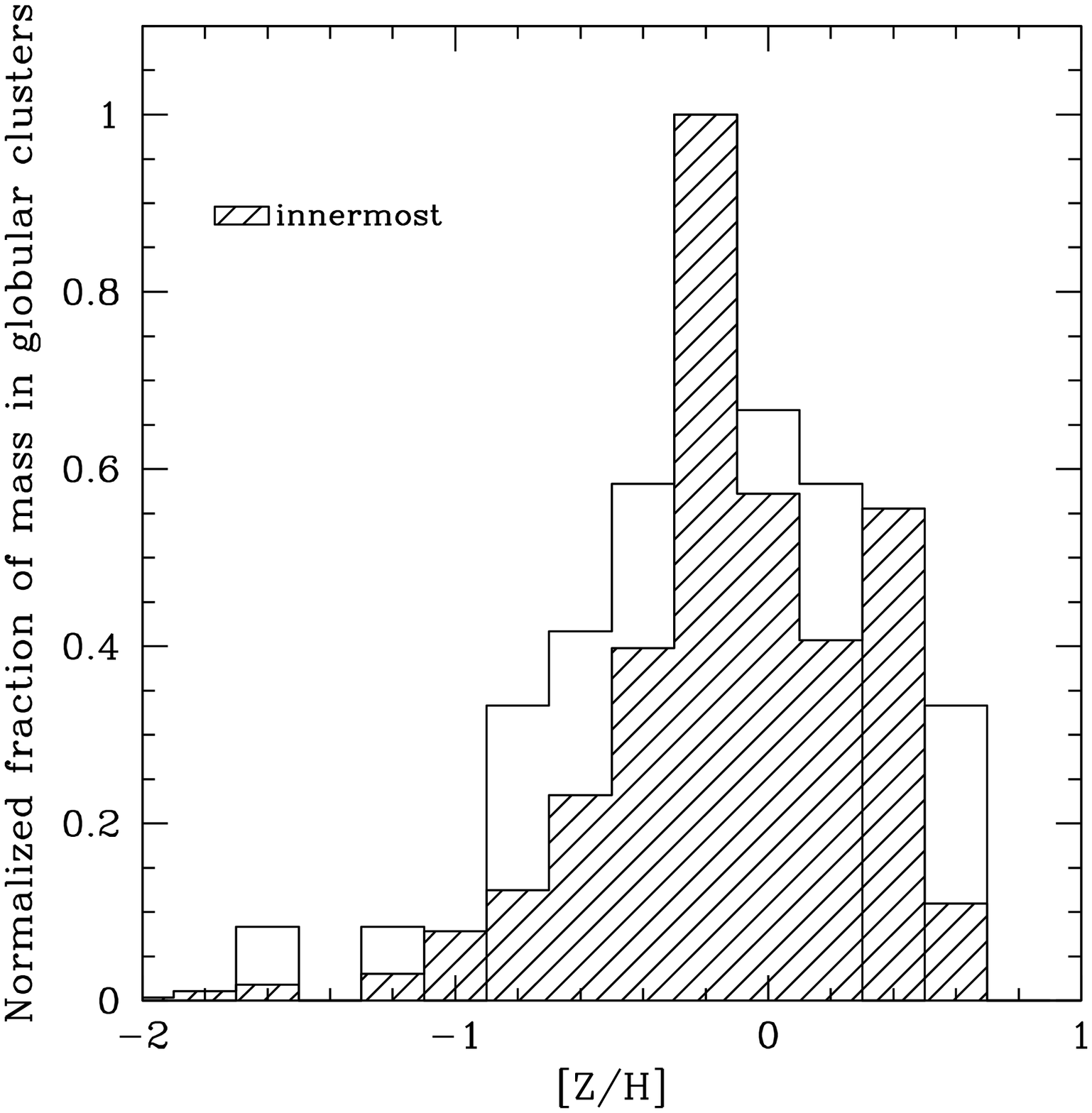}{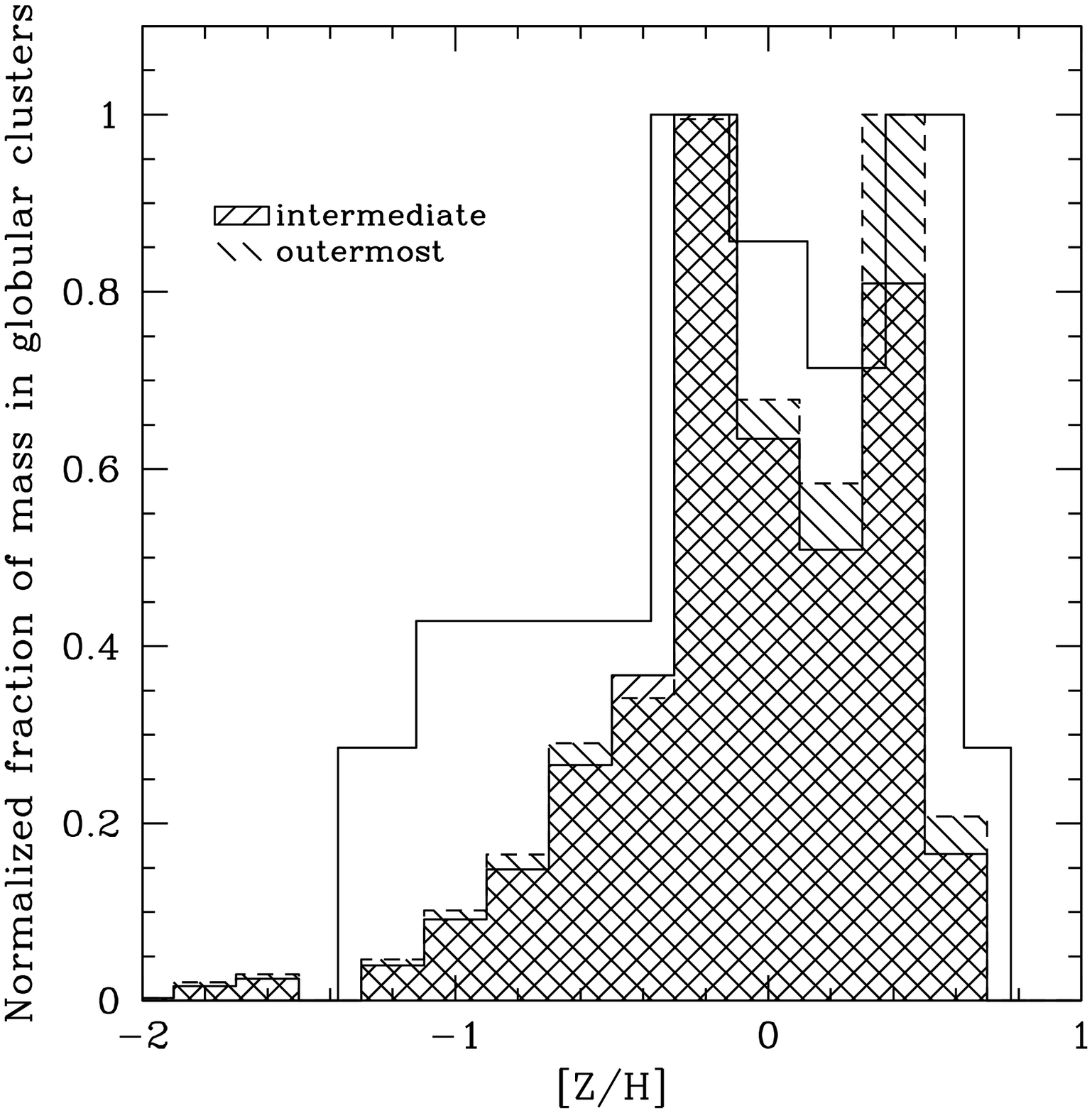}

\caption{Predicted globular-cluster metallicity distribution $\Upsilon_{\rm GC,
tot}$ by mass as a function of [Z/H] for three different radial
compositions (i.e. $f_{red}/f_{blue}$). The left panel shows both model
predictions and observations related to the central part of an elliptical
galaxy. The right panel shows the same quantities for cluster populations
residing at $r \ge R_{\rm eff}$. Solid empty histograms:
observational data taken as sub-samples of the P06 compilation, according
to the galactic regions presented in each panel.}

\label{GD_mass}

\end{figure*}

\begin{figure*}


\plottwo{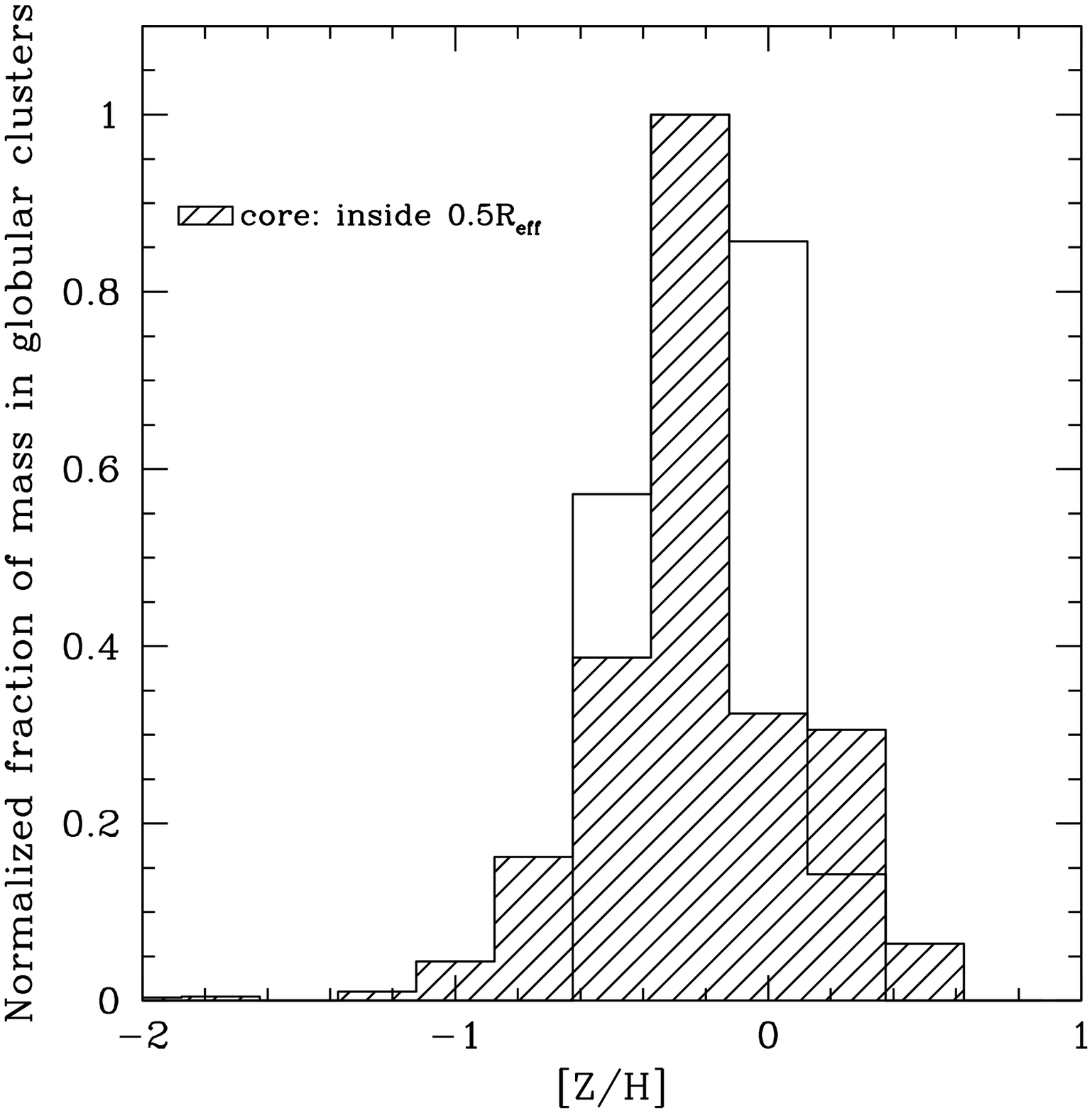}{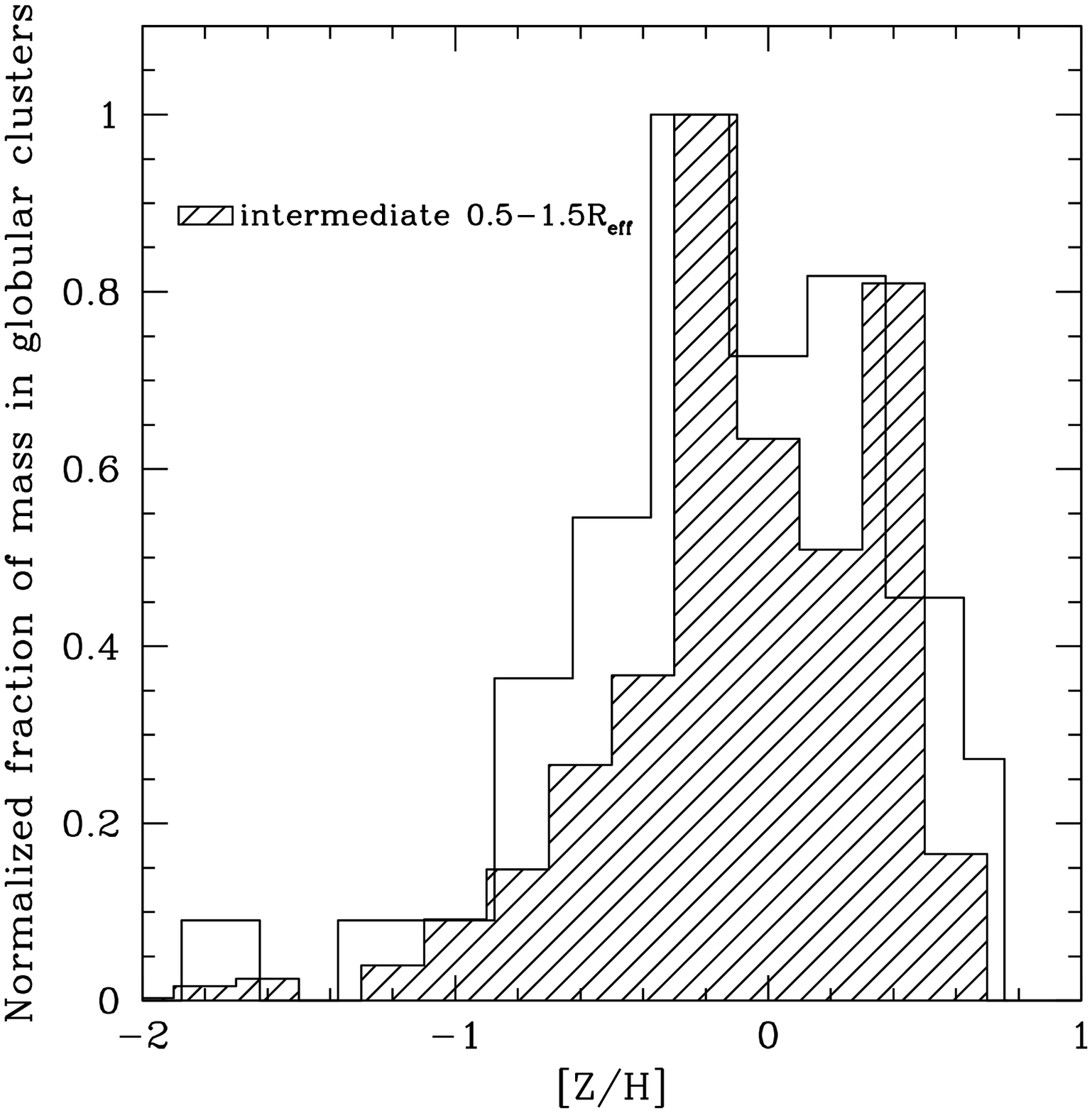}

\caption{Predicted globular-cluster metallicity distribution $\Upsilon_{\rm GC,
tot}$ by mass as a function of [Z/H] for two different projected
galactocentric radii. The left panel shows both model predictions and
observations related to the \emph{pure core} of an elliptical galaxy
(namely $f_{red}:f_{blue}=1:0$). The right panel shows the same quantities
for cluster populations residing either at $0.5R_{\rm eff} < r <1.5 R_{\rm
eff}$. Solid empty histograms:
observational data taken as sub-samples of the P06 compilation, according
to the galactic regions presented in each panel.}
\label{GD_mass2}

\end{figure*}

\subsubsection{The Comparison Sample}

\label{compsample}

As stressed in the introduction, the multi-modality in GCSs varies as a
function of host galaxy properties (e.g.~mass, morphological type, etc.).
Here we try to match the distributions resulting from the recent
compilation of spectroscopic data by Puzia et al.~(2006, hereafter P06),
which samples the typical bimodal color distribution of GCs in nearby
galaxies \citep[see also][]{puzia04, puzia05}. This is illustrated in
Figure~\ref{colsamp}, where we plot the $(V\!-\!I)_0$ color distribution
of the P06 sample of GCs in elliptical galaxies ({\it top panel}) together
with those of GCs in NGC~4472 and the Milky Way ({\it middle and bottom
panel}). NGC~4472 is the most luminous elliptical in the Virgo galaxy
cluster and hosts a GC system with a prototypical color bimodality
\citep[e.g.][]{puzia99}. To allow direct comparison with the P06 sample,
we use GCs in NGC~4472 that are brighter than $V\simeq22.5$ since the P06
sample includes only the brightest GCs in nearby early-type galaxies in
order to maximize the S/N of their spectra. The resulting color
distribution is remarkably similar to the one of the P06 sample, which
assures {that the P06 sample includes} a representative sampling of
the GC color bimodality in massive elliptical galaxies.

However, the comparison with the Milky Way GCs shows that the P06 sample
covers few of the most metal-poor GCs. Therefore, the \emph{bimodality}
which we refer to in the following may not be the same as the one observed
in spirals or in some elliptical {galaxies}, where a substantial population of
metal-poor clusters with [Z/H]~$\la -1.5$ is present \citep[e.g.][]{gk99}.

{We do not include the dynamical evolution of GCs in our model, since we are considering only the brightest (most massive, i.e. $>10^{5.5} M_{\odot}$, see also Puzia et al.~2004, A\&A 415, 123) clusters in nearby galaxies as comparison sample. The comparison sample includes GCs much brighter than the typical turnover magnitude of the globular cluster luminosity function and we, therefore, do not expect significant differential dynamical evolution for these massive systems \citep[see][for details]{gnedin97}. In fact, it has been shown (e.g. Fall \& Zhang, 2001) that the timescales for both evaporation by two body relaxation and tidal stripping of star clusters is longer than a Hubble time for GCs more massive than $\sim 10^{5.5}\, M_{\odot}$.} 

In our model, the number of clusters formed in a star-formation burst of a given
strength is adjusted to match the observations. Hence, the absolute
scaling of GC numbers is arbitrary, i.e., within physical limitations of
the star formation rate any number of GCs can be reproduced by adjusting
the function $f$ in Equation~\ref{gcmd1}.

If, however, metal-poor and metal-rich GCs are on systematically different
orbits and experience significantly different dynamical evolutions the
effect {of tidal disruption} might be slowly changing relative GC
numbers with time. Another complication is the variation of the initial
star-cluster mass function, in particular as a function of metallicity.
Modeling these effects requires detailed knowledge of the orbital
characteristics and chemo-dynamical processes that lead to star
cluster formation, and goes far beyond the scope of this work. {We keep
these potential systematics in mind, but expect negligible impact on our analysis.}

\subsubsection{A Simple Model}

In order to make a first-order comparison between our model predictions
and the observed $\Upsilon_{\rm GC, tot}$ at $t\!=\!13$ Gyr, we first
focus on the simple case in which:

\begin{eqnarray}
\Upsilon_{\rm GC, tot}(Z)=
f_{red} \cdot \Upsilon_*(t=13\!~{\rm Gyr},r_1,Z) + \nonumber\\
f_{blue} \cdot \Upsilon_*(t=13\!~{\rm Gyr},r_2,Z) \; ,
\label{gcmd2}
\end{eqnarray}

with $f_{red}\, , f_{blue}=const$ and $r_1 = 0.1 R_{\rm eff}$, $r_2 \ge 1
R_{\rm eff}$. The first term ($red$) corresponds to a population
typical of the galaxy core (well inside  $r < 1 R_{\rm eff}$). The second
term ($blue$) represents $\Upsilon_*$ in the outer regions. In order to
take into account the different amounts of stars formed in each galactic
region, we point out that the stellar metallicity distributions entering
Equation~\ref{gcmd1} were not normalized.

As a first step, we take several values for the weights $f_{red} \, , f_{blue}$ in
order to mimic different mixtures of the two GC populations.
In particular, we used the relative numbers of the \emph{red} (here
identified as the metal-rich core population) and the \emph{blue} globular
clusters (the \emph{halo} metal-poor population) as a function of
galactocentric radius for the elliptical galaxy NGC~1399 \citep{dirsch03}.
Our particular choice is driven by observationally motivated values for
the weights, although we realize that NGC~1399 is a quite peculiar,
massive cD elliptical and might not be representative of less massive

systems. In the context of this first step, the weights might reflect the
effects of the projection on the sky of a three-dimensional structure.
However, we show below that the results do not strongly depend on the
weights. Therefore they might be interpreted as mean values and could be
changed if one decides to model a particular galaxy, {with a different ellipticity,
inclination, and luminosity profile.}

In Figure~\ref{GD_mass} we show the globular-cluster metallicity
distribution $\Upsilon_{\rm GC}$ by mass (computation based on
$\Upsilon_{*,m}$) in two radial bins for three particular choices of
weights. In particular in the following {we} will use the ratios
$f_{red}\!=0.77 ,\!f_{blue}=0.23$, $f_{red}\!=0.60,\!f_{blue}=0.40$ and
$f_{red}\!=\!f_{blue}=0.50$ in order to define the theoretical
\emph{innermost, intermediate and outermost} sub-sample of the GCS,
respectively. These $\Upsilon_{\rm GC}$ will be compared against
subsamples of the {P06 data}, obtained by selecting GCs with either $r < 1
R_{\rm eff}$ (in the case of the innermost population) or $r \ge 1 R_{\rm
eff}$ (for the intermediate and the outemost cases, respectively), unless
otherwise stated.

\subsubsection{Globular Cluster Metallicity Distribution}

In order to plot the different cases on the same scale we normalize each
$\Upsilon_{\rm GC}$ by its maximum value. In the left panel of
Figure~\ref{GD_mass} the shaded histogram represents the innermost
population. Our predictions match the data very well, especially in the
metal-rich slope and the mean of the distribution. The same happens for
the \emph{pure core} populations, which shows how the GCS might be used to
probe the CSP in ellipticals. It should be remarked that a second peak
centered at super-solar metallicity appears in the distribution predicted
by our models, although not evident in the data of the particular radial
sub-sample. The right panel of Figure~\ref{GD_mass} illustrates model
predictions which are more representative of the galaxy as a whole (either
at $1 R_{\rm eff}$, i.e. the \emph{intermediate} population, or at several
effective radii, the \emph{outermost} population), and we consider them as
the fiducial case. These two cases look quite similar to each other and
have clear signs of bimodality in remarkable agreement with the
spectroscopic data (solid empty histogram, sub-sample of the P06 data with
$r \ge R_{\rm eff}$). { A {Kolmogorov-Smirnov} test returns $>\!99\%$
probability that both model predictions and observations are drawn from
the same parent distribution in the left panel of Figure~\ref{GD_mass}.
The right panel statistics gives a lower likelihood of $98.4\%$ that both
distributions have the same origin, which is mainly due to the observed
excess of metal-poor GCs at large galactocentric radii compared to the
model predictions.} The prediction of a super-solar metallicity globular
cluster sub-population is entirely new and a result of the radially
varying and violent formation of the parent galaxy. Moving to the
low-metallicity tail, we predict slightly fewer low metallicity objects
than expected from observations. But we recall the systematics mentioned
in Section~\ref{compsample}.

In Figure~\ref{GD_mass2} (left panel) we show the results for a \emph{pure
core} GCs, namely one in which we adopt $f_{red}\!:\!f_{blue}=1\!:\!0$. In
this quite extreme case the observed GCs have been selected with radius $r
< 0.5 R_{\rm eff}$. The histogram reflects the shape of a G-dwarf-like
diagram expected for a typical CSP inhabiting the galactic core. This
finding is particularly important, because it might offer the opportunity
to resolve the SSPs in ellipticals, at variance with data coming from the
integrated spectra which deal with luminosity-weighted quantities. Whereas
in Figure~\ref{GD_mass2} (right panel), the intermediate population is
compared to a sub-sample of P06 GCs with $0.5 < r < 0.5 R_{\rm eff}$. This
is to show that the multimodality is not an artifact due to the particular
radial binning adopted in this paper.

\begin{figure}


\plotone{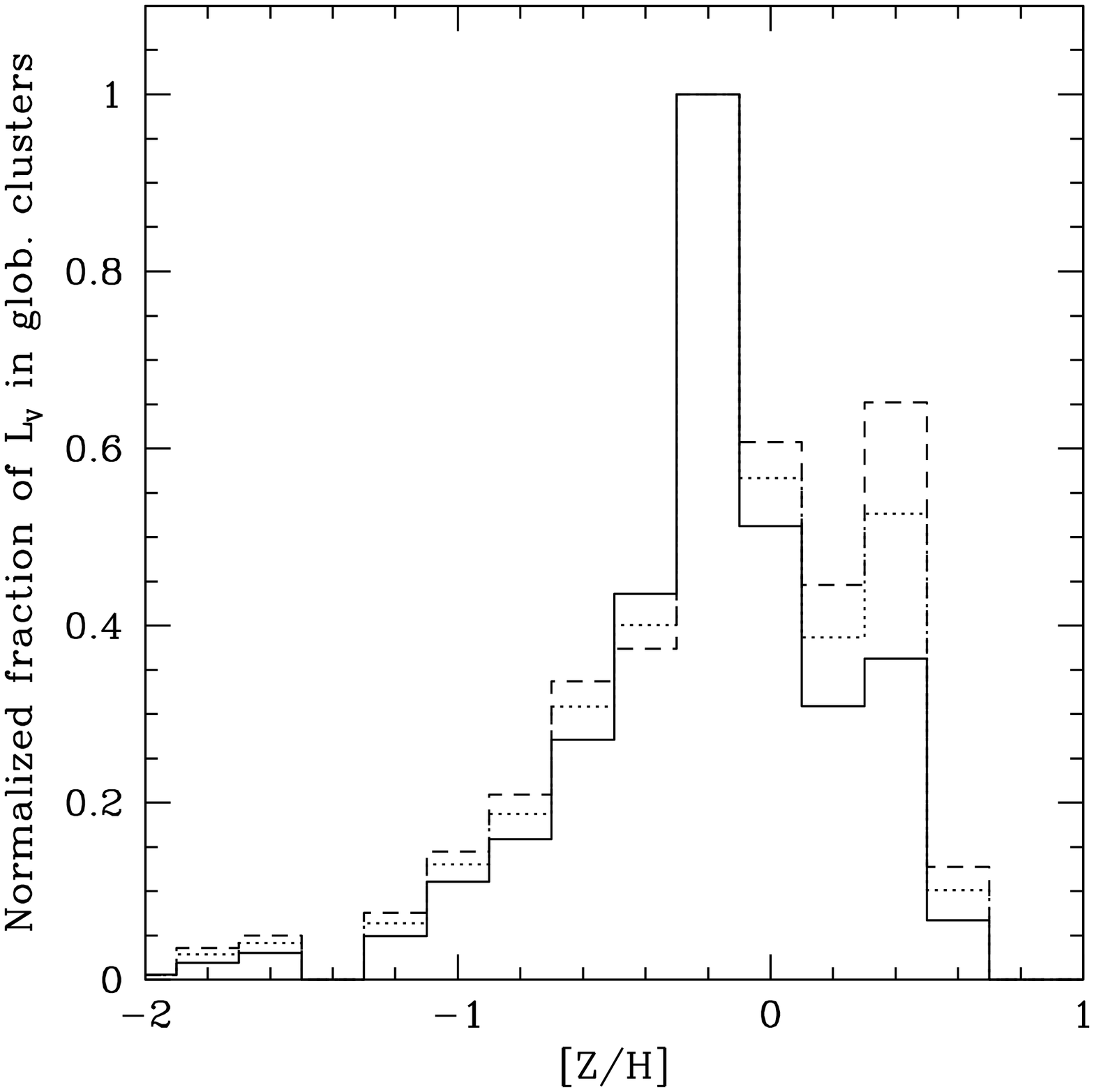}

\caption{Predicted globular-cluster metallicity distribution $\Upsilon_{\rm GC,
tot}$ by luminosity at three different projected galactocentric radii.
Solid: innermost region; dotted: average galactic (intermediate
population); dashed: outermost part.}
\label{GD_lum}

\end{figure}

Figure~\ref{GD_lum} shows the $V$-band luminosity-weighted $\Upsilon_{\rm
GC}$ for which the computation is based on $\Upsilon_{*,l}$. This
metallicity distribution has been obtained by converting the mass in each
[Z/H] bin of the previous figure into $L_V$, by means of the $M/L_V$ ratio
computed by \cite{maraston05} as a function of [Z/H] for 13 Gyr old SSPs.
Due to the well-known increase of the $M/L_V$ in the high metallicity tail
of the distribution\footnote{See PMC06 for a comparison between G-dwarf
like diagrams for $\Upsilon_{*,m}$ and $\Upsilon_{*,l}$ predicted for the
same CSP.}, we notice in Figure~\ref{GD_lum} that now the second peak has
a smaller intensity in all the cases. {The corresponding diffuse-light
population goes undetected in integrated-light studies.} In any case, the
conclusions reached by analyzing Figure~\ref{GD_mass} are not
significantly altered. We conclude that our analysis is not significantly
biased by some metallicity effect which may alter the shape of the
observed $\Upsilon_{\rm GC, tot}$ by luminosity. We stress the power of
GCSs in disentangling stellar sub-populations in massive ellipticals, due
to their nature as simple stellar populations that can be directly
compared to SSP model predictions, unlike diffuse light {measurements}.

Even with this simple parametrization, where $f$ in Equation~\ref{gcmd2}
does not depend on metallicity, we suggest that the bimodality for the
metal-rich GCs is the result of different shapes of
$\Upsilon_{*,m}$ (and the $\Upsilon_{*,l}$) at different galactocentric
radii.

\begin{figure*}


\plottwo{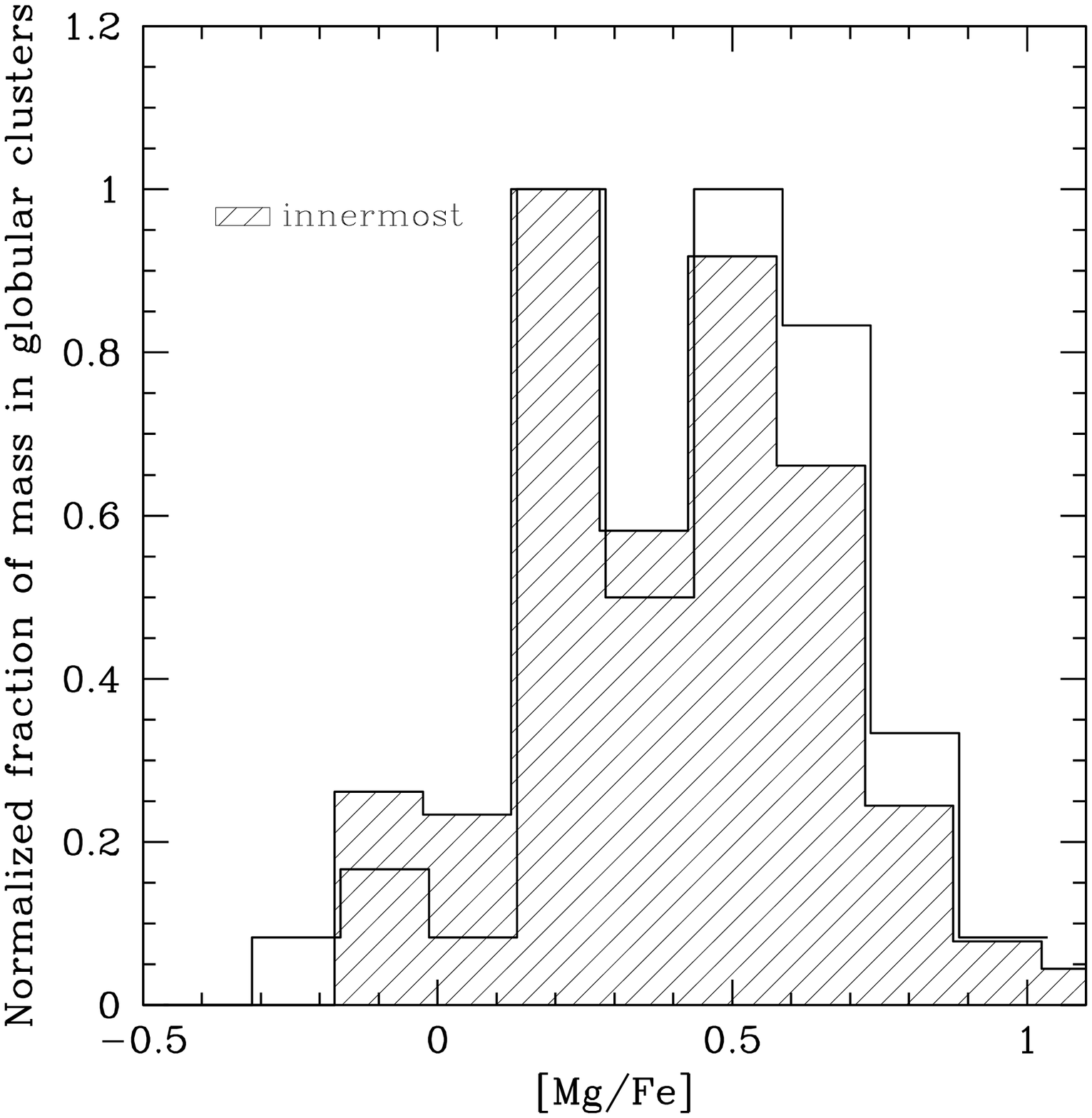}{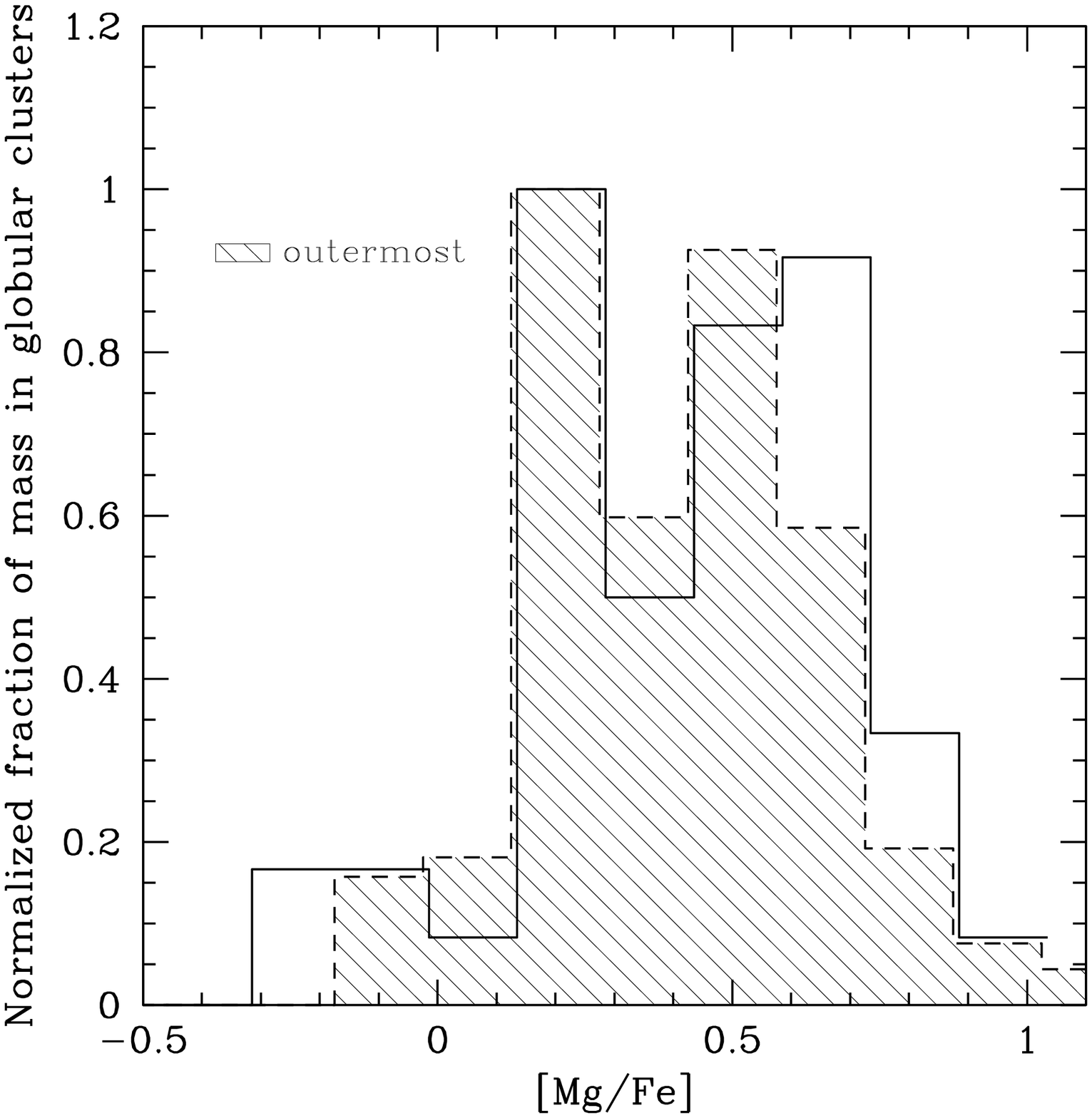}

\caption{Shaded histogram: Predicted distribution of globular-cluster [Mg/Fe]
values at two different projected galactocentric radii (innermost and
outermost regions). Solid empty histogram: observational data taken as
sub-samples of the P06 compilation, according to the galactic regions
presented in each panel (see text).}
\label{GD_mgfe}

\end{figure*}

\subsubsection{Globular Cluster [Mg/Fe] Distributions}

Finally, in Figure~\ref{GD_mgfe} we show the [Mg/Fe] distributions for GCs
divided in radial bins as in Figure~\ref{GD_mass}. According to PMC06,
these [Mg/Fe] distributions are narrower, more symmetric, and exhibit a
smaller radial variation with respect to the [Z/H] distributions. In any
case, a small degree of bimodality is still present. We point out the
impressive agreement with the spectroscopic observations by P06. 
There is a rather large discrepancy between data and models at the high-[Mg/Fe] end.  
Hence, the corresponding {Kolmogorov-Smirnov} likelihood tests return
a probaility of 1\% (for inner sample) and 95\% (for the outer ones).
If we limit the model predictions to [Mg/Fe] $<$ 0.8 dex, the agreement slightly improves, reaching a 10\% probablity in the inner region.
However, the inner field data still does not reach the extreme [Mg/Fe] values of the models. 
In fact, due to the monotonic decrease of the [Mg/Fe] as a function of
either metallicity or time (see PM04), the lack of low-metallicity GCs,
evident from Figure~\ref{GD_mass}, translates into a lack of
$\alpha$-enhanced clusters.

{ The [Mg/Fe] bimodality of our model predictions and the match with the
spectroscopic measurements strongly imply that globular clusters in
massive elliptical galaxies form on two different timescales.} Their
chemical compositions are consistent with an {\it early} mode with a duration of
$\Delta t\la\ 100$ Myr and a \emph{normal} formation that lasted for $\Delta
t\la 500$ Myr. 
{In fact, according to the time-delay model (see Matteucci, 2001)
and given the typical star formation history of our model ellipticals,
the [Mg/Fe] ratio in the gas - out of which the GCs form - is
quickly and continuously decreasing with time. 
We predict that the [Mg/Fe] ratio can be higher than 0.65 dex (i.e. in the bins
in which our predictions exhibit a deficit of GCs with respect 
to the observed distribution) only in the first $\sim 100$ Myr   
(see also Fig.3 in P06 and related discussion).
In fact, such a high value for the [Mg/Fe] can be attained only
if very massive type II SNe contribute to the chemical evolution,
without any contribution from either lower-mass type II or type Ia SNe.
The \emph{normal} formation, instead, is the one already plotted in Fig.~\ref{GD_mgfe}
and forms on a typical timescale of $0.5-0.7$ Gyr.
More quantitatively, our initial theoretical GCMD predicts that only $\sim 4\%$
of the GCS forms at [Mg/Fe] larger than 0.65 dex. 
In order to improve the agreement with observations we require
that the above fraction should be increased to $\sim 12-15\%$.}
Since star and globular cluster formation are expected to
be closely linked \citep[e.g.][]{chandar06} the same must be true for the diffuse
stellar population of such galaxies. We therefore foresee the presence of
a similar [Mg/Fe] bimodality in the diffuse light of massive elliptical
galaxies.
Unfortunately, there are not direct observations confirming our
suggestions, until the {metallicity distributions} 
for the diffuse {stellar component} in ellipticals will become available for
a number of galaxies. Indeed, the detection of 
bimodality in the [Mg/Fe]-distribution might be a benchmark 
{test} for our predictions.



We will still refer to multiple GC sub-populations. However, their
differences { ought} to be ascribed only to the fact that they are
created during an extended (and intense) star formation event during which
the {variation in} chemical evolution is not negligible. The radial
differences originate from the fact that the galactic wind epoch is
tightly linked to the potential, occurring later in the innermost regions
\citep[e.g.][]{carollo93, martinelli98}. Moreover, PM04 and PMC06 found
that also the infall timescale is linked to the galactocentric radius. In
particular, it lasts longer in the more internal regions, owing to the
continuous gas flows in the center of the galactic potential well. In
particular, we recall that in our model the core experiences a longer
($\sim\!0.7$ Gyr) star-formation history with respect to the outskirts
where the typical star-formation timescale is $\sim\!0.2$ Gyr.

According to our models the metal-rich population of GCs in massive
elliptical galaxies may consist of multiple sub-populations which
basically play the same role as the CSPs populating each galactocentric
shell in our framework of the global galaxy evolution. {At the same
time, we point out the lack of our models to produce a significant
fraction of metal-poor GCs similar to the halo GC population in the Milky
Way, with the caveat that the star formation histories are very different. 
This, in turn, suggests that GCSs in giant elliptical galaxies were
assembled by accretion of a significant number of metal-poor GCs.}


\subsection{Metallicity Dependent Globular Cluster Formation}

\label{ref:metals}

\begin{figure*}



\plottwo{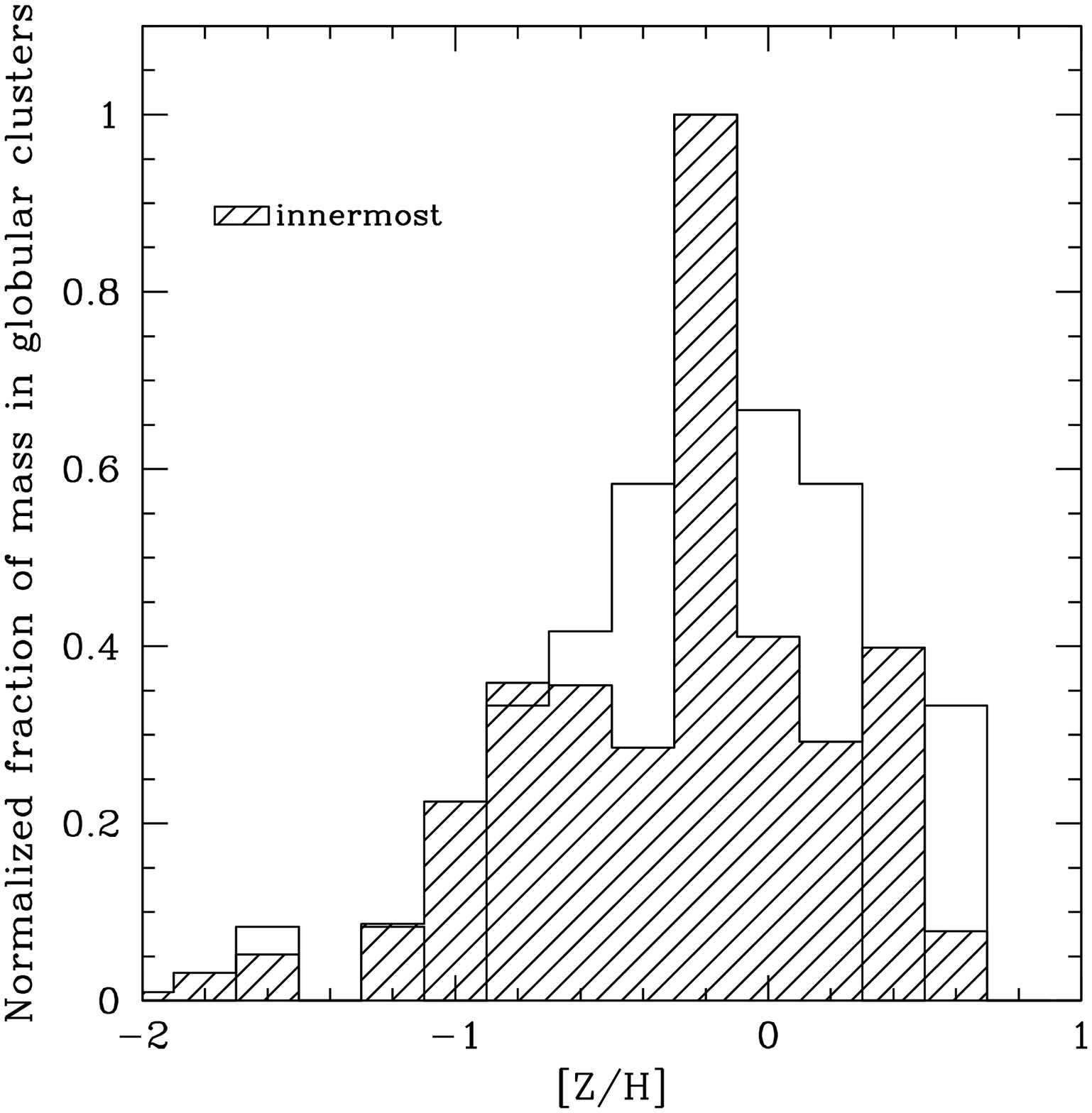}{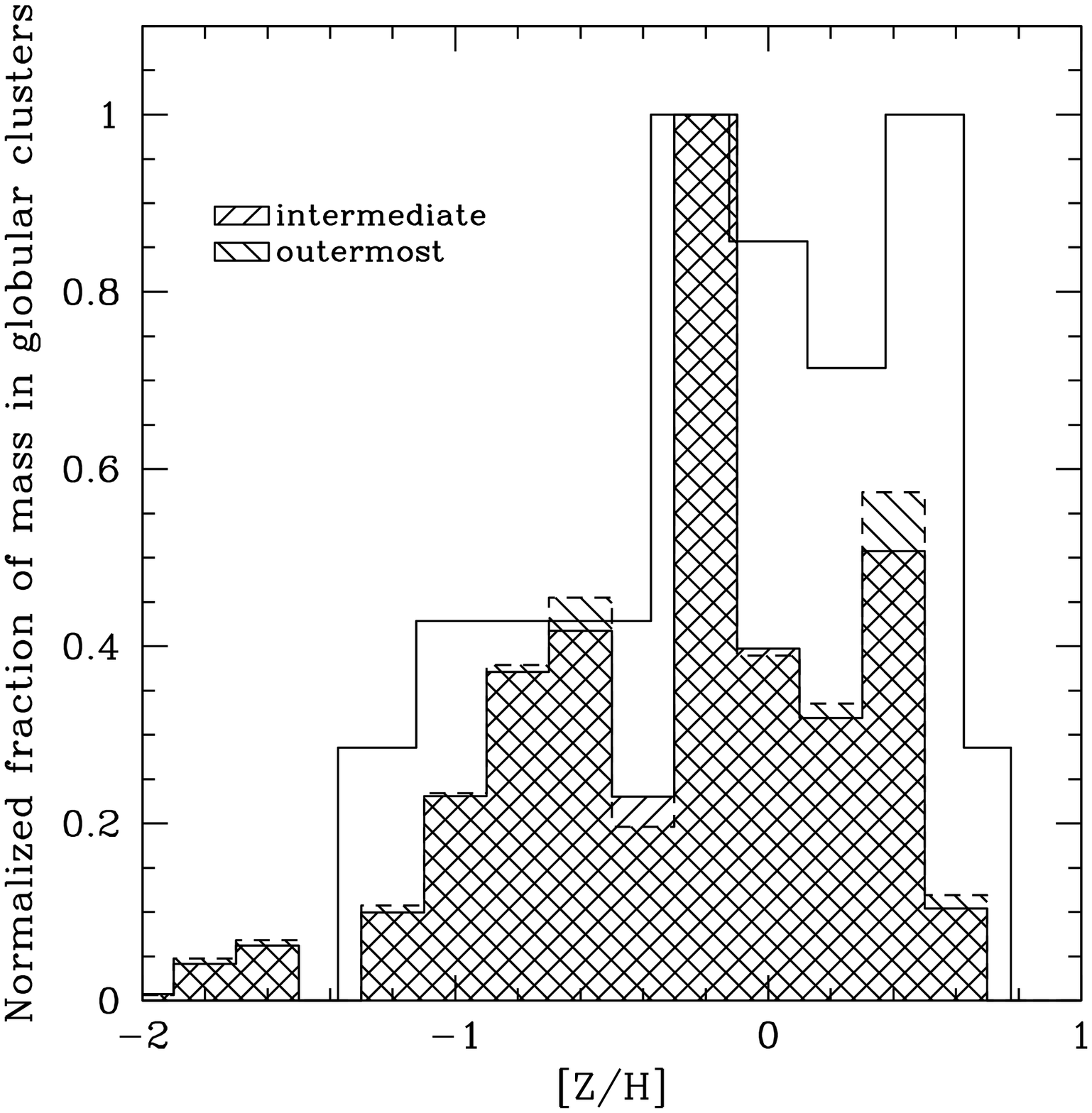}

\caption{Predicted globular-cluster metallicity distribution $\Upsilon_{\rm GC,
tot}$ by mass as a function of [Z/H] for three different radial compositions 
(i.e. different $f_{red}/f_{blue}$). In this case, the function \emph{f} has
an explicit dependence on [Z/H] (see text). The left panel shows both
model predictions and observations related to the central part of an
elliptical galaxy. The right panel shows the same quantities for cluster
populations residing at $r \ge R_{\rm eff}$. Solid empty histograms:
observational data taken as sub-samples of the P06 compilation, according
to the galactic regions presented in each panel.} 
\label{GC_HH}

\end{figure*}

In this section we explore the approach outlined in Equation~\ref{gcmd1},
by introducing the effect of metallicity in the function \emph{f}. In
particular, we start by assuming that
\begin{equation}
\label{metaldepnd}
\frac{f(t,r_i,{\rm [Z/H]}\!<\!-1)}{f(t,r_i,{\rm [Z/H]}\!>\!-1)}=2 \; ,
\end{equation}
roughly following what was found {for the GCS of the most nearby giant
elliptical galaxy NGC~5128 (Centaurus A)} by \citeauthor{harris02}
(\citeyear{harris02}, hereafter HH02) for their ``inner field'',
regardless of the radius of the $i$-th shell. Since the final
distributions are normalized, the actual zeropoint of the function $f$ is
not relevant. Our model predictions are plotted in Figure~\ref{GC_HH}. We
notice a modest increase of the low-metalliticy tail of the distribution
with respect to the simple picture sketched in Section~\ref{bimod} without
metallicity dependence, as well as a lower fraction of globular clusters
populating the high-metallicity peak. {Including the metallicity
dependence leads to an ambiguous change in agreement with the general
observed trend. Hence,} no firm conclusions on the real need for a
metallicity dependence can be drawn. Similar results are obtained in the
more realistic case in which \emph{f} is a linearly decreasing function of
[Z/H].

At variance with HH02, we chose to adopt the same scaling irrespective of
galactocentric radius, for the following reason. Despite the fact that the
HH02 stellar metallicity distributions $\Upsilon_*$ as functions of [Z/H]
confirm both the shape and the radial behavior of our model predictions
for the mass-weighted stellar metallicity distribution $\Upsilon_{*,m}$
(compare their Fig.~7 with PMC06 Fig.~4), care should be taken when
comparing their results for $\Upsilon_*$ as a function of [Fe/H]. The
latter, in fact, had been obtained by assuming a particular trend in the
[$\alpha$/Fe] { as a function of galactocentric radius} which disagrees
with the results of our detailed chemical evolution model. In particular,
we find an offset of at least 0.2 dex in the sense that [Fe/H]~$_{\rm
HH02}\!\sim\! 0.2 +$~[Fe/H]~$_{\rm PM04}$ at a given metallicity ([Z/H]).
{This} disagreement becomes larger either at very low metallicity or
at larger galactocentric radii, where we expect a stronger
$\alpha$-enhancement. Once the PM04 value for [Fe/H] is adopted in Fig.~18
of HH02, we find that: $i)$ for the inner halo, the stellar
$\Upsilon_{*,m}$ should be shifted by $\sim\!0.2$ dex toward lower
metallicities, removing any metallicity effect, and $ii)$ for the outer
halo the discrepancy between the stellar $\Upsilon_{*,m}$ and the
$\Upsilon_{\rm GC, tot}$ should be reduced.

Nevertheless, we believe that some decrease with time of the function
\emph{f} could be motivated by theoretical arguments. In fact, recent work
\citep[e.g.][]{elmegreen97, elmg04} shows that GCs of all
ages preferentially form in turbulent high-pressure regions. If we
interpret the decrease in the efficiency of star formation (inside the gas
clouds that form GCs), as a function of the ambient pressure
\citep{elmegreen97} as a proxy for the temporal behaviour of our function
\emph{f}, we find again a reduction of a factor $\sim\!2\!-\!3$ from the
early high-pressure epochs to a late, more quiescent evolutionary phase.

\subsubsection{The Ratio of Metal-poor to Metal-rich Globular Clusters}

For our fiducial model we predict a ratio of metal-poor (namely with
[Z/H]~$\le-1$) to metal-rich GCs of $\sim\!0.2$. Previous
photometric surveys found that the typical value for GC
systems in elliptical galaxies is close to unity \citep{gk99, kundu01a}.
Provided a linear color-metallicity transformation (see also
Section~\ref{othermechs}), a possible explanation for the discrepancy
between our models and the observations might be 
obtained by boosting the metal-poor
population by a factor of ${f(t,r_i,{\rm [Z/H]}\!<\!-1)}/{f(t,r_i,{\rm
[Z/H]}\!>\!-1)}\ge 5$.

Another way to solve the discrepancy is to assume that all the
\emph{missing} {globular clusters} have been accreted from the
surroundings, e.g. from dwarf satellites \citep[e.g.][]{cote98}. We
estimate the amount of the accreted metal-poor GCs, needed to achieve a
ratio close to 1, as a factor of $\sim\!4$ of the number of {globular
clusters initially} formed inside the galaxy.


\subsection{The Role of the Host Galaxy Mass}

A natural consequence of the scenario depicted in Sections~\ref{bimod} and
\ref{ref:metals} is that, at a given galactocentric radius, the mean
metallicity and [$\alpha$/Fe] ratios of a GCS coincide with the
mass-weighted [$\langle$Z/H$\rangle_*$] and
[$\langle\alpha$/Fe$\rangle_*$] of the underlying stellar population,
because the GC quantities are calculated either from $\Upsilon_{*,m}$ or
the $\Upsilon_{*,l}$ (see Eqs.~1 and 2 of PMC06), unless the scaling
function \emph{f} is allowed to strongly vary with time. We expect this to
happen at least in the innermost GC sub-populations, in
which the effects of the accretion of GCs from the
environment can be reasonably neglected. In particular, PM04 predict that
more massive galaxies should show higher [$\langle\alpha$/Fe$\rangle_*$]
and [$\langle$Z/H$\rangle_*$]. If accretion plays a negligible role, we
expect the same correlations for the {\it total} GC
population with host galaxy mass for the most massive systems, in
agreement with current observations \citep[e.g.][]{vdb75, brodie91,
peng06}.

In fact, if we perform the same study of the above sections for a $10^{12}
M_{\odot}$ galaxy (see Table 2 of PM04 for its properties), both peaks in
$\Upsilon_{\rm GC, tot}$ shift their positions by about 0.2 dex to higher
[Z/H]. This is in good agreement with the results of \cite[see their
Figure~13 and 14]{peng06}. This trend holds for smaller objects as well.
If we apply the procedure to a $10^{10} M_{\odot}$ galaxy (Model IIb of
PM04), we find that the metal-rich peak shifts towards a lower metallicity
by 0.3 dex (with respect to our fiducial model with $M_{\rm lum}=10^{11}
M_{\odot}$), while the other peak is now centered around [Z/H]~$=-0.8$
dex. In particular, we find a faster decrease in the mean metalliticy of
the metal-poor GCs than for the metal-rich ones, again in agreement with
the \citeauthor{peng06} results.

Interestingly, the ratio of metal-poor to metal-rich cluster increases up
to $\sim\!0.5$ {for less massive halos}. We recall that in the PM04
scenario, the low-mass galaxies are those forming on a longer timescales
and with a slower infall rate. Therefore, we suggest that the combination
of these factors is likely to at least partly explain the change of the GC
distributions in different galaxy morphologies. This is especially the
case in dwarf galaxies, where star formation is slow and still on-going,
together with the fact that the probability for a substantial change in
the pressure of the interstellar medium relative to its initial values is
higher than in ellipticals, thus implying a much stronger variation of $f$
with time.


\subsection{Merger-Induced Globular Cluster Formation}

\label{ln:merger}

It has been suggested that GC populations are produced
during major merger events which would lead to present-day ellipticals and
their rich GCSs \citep[e.g.][]{schweizer87, ashman92}. Subsequent studies
\citep[e.g.][]{forbes97, kp98b} challenged this view by pointing out the much
higher $S_N$ and more metal-rich GCSs in early-type
galaxies compared to those of spiral and irregular galaxies, which are
thought to represent the early building blocks of massive ellipticals.

In the following, we study the impact of the merger hypothesis on the
predictions of our simulations. In order to do that, we extended the
procedure sketched in the previous sections to the \emph{merger models}
presented by \citeauthor{pipino06b} (\citeyear{pipino06b}, hereafter
PM06). In this paper, the effects of late gas accretion episodes and
subsequent merger-induced starbursts on the photo-chemical evolution of
elliptical galaxies have been studied and compared to the picture of
galaxy formation emerging from PM04; in particular the PM04 best model is
taken here as a reference. By means of the comparison with the
colour-magnitude relations and the $[\langle$Mg/Fe$\rangle_V]$-$\sigma$
relation observed in ellipticals \citep[e.g.][]{renzini06}, PM06 showed
that either bursts involving a gas mass comparable to the mass already
transformed into stars during the first episode of star formation and
occurring at any redshift (major mergers), or bursts occurring at low
redshift (i.e.~$z\le0.2$) and with a large range of accreted mass (minor
mergers), are ruled out. The reason lies in the fact that the chemical
abundances in the ISM after the galactic wind (and before the occurrence
of the merger) are dominated by Type Ia SN explosions, which continuously
enrich the gas with their ejecta (mainly Fe). When the merger-induced
starburst occurs, most stars form out of this enriched gas (thus, e.g.,
lowering the total $[\langle$Mg/Fe$\rangle]$); at the same time, we expect
the metallicities of GCs formed out of this gas to be on average higher
{ and their [Mg/Fe] ratios to be lower} than those of the bulk of stars and
GCs formed in the initial
starburst.

\begin{figure}[!t]


\plotone{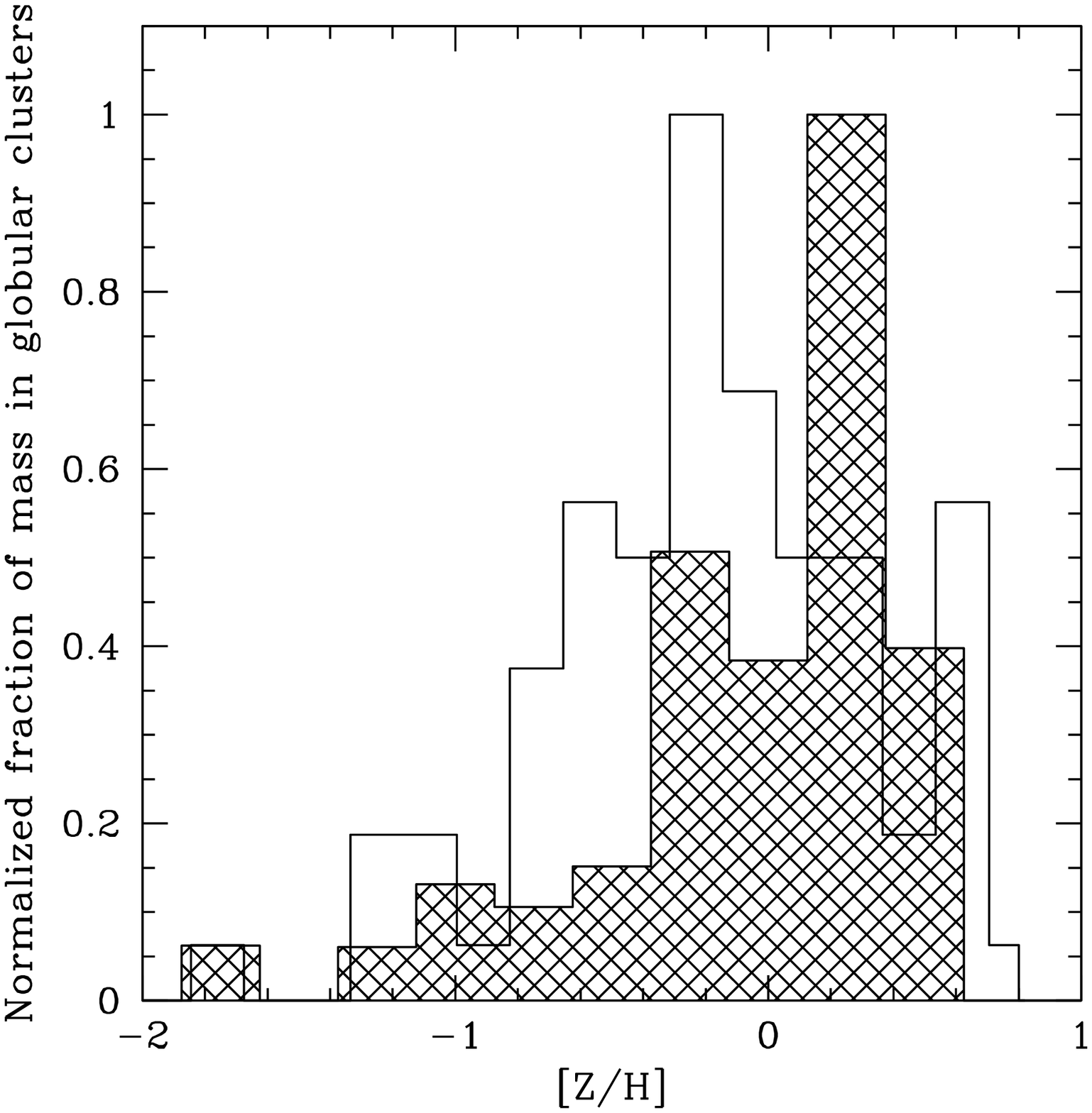}

\caption{Shaded histogram: predicted total GC metallicity
distribution $\Upsilon_{\rm GC, tot}$ by mass for the $<1 R_{\rm eff}$ shell,
for a case in which a second episode of star formation, induced by a
gaseous merger, is taken into account (see text). Solid histogram:
observations from \cite{puzia06}, their entire sample.} 
\label{GD_merg}

\end{figure}

In this work we present the case in which the galaxy accretes a gas mass
$M_{acc}=M_{\rm lum}$ at $t_{\rm acc}=2$ Gyr (i.e.~$\sim\!1$ Gyr after the
onset of the galactic wind). We make this choice for several reasons: $i$
This model is quite similar to the PM06 models \emph{b} and \emph{g},
which were among those in good agreement with observations of the diffuse
galaxy light properties. $ii$) The formation epoch of the bulk of these
\emph{second generation} GCs cannot occur $\ga\!2$ Gyr later than $t_{\rm
gw}$, because the majority of GCs in the most massive elliptical galaxies
studied today appear old within the age resolution of current photometric
($\Delta t/t\approx0.4\!-\!0.5$) and spectroscopic studies ($\Delta
t/t\approx0.2\!-\!0.3$). Finally, the composition of the newly accreted
gas is assumed to be primordial (see PM06 for a detailed discussion), but
we remark that we reach roughly the same conclusion in the case of solar
composition, in order to mimic some pre-enrichment for the newly accreted
gas. We point out that, lacking dynamics, PM06 presented their results for
one-zone models. Therefore, in this section we are considering
Equation~\ref{gcmd1} limited to only one shell. In this way we can check
whether the single merger hypothesis alone is enough to produce some
bimodality in the total globular-cluster metallicity distribution
$\Upsilon_{\rm GC, tot}$, and if it is consistent with the predictions
based on our fiducial model described in Section~\ref{ref:metals}.

\begin{figure}[!t]


\plotone{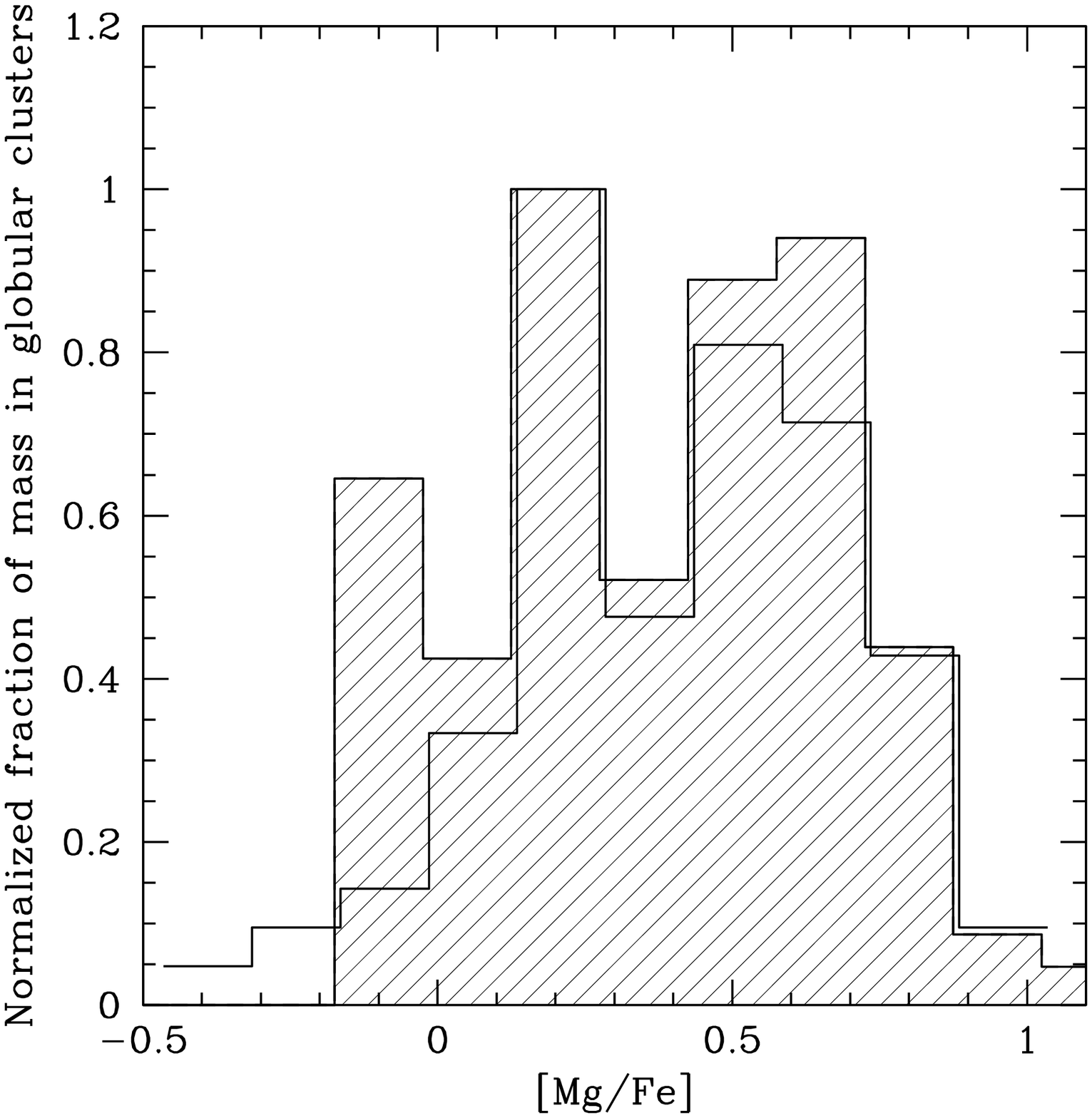}

\caption{Shaded histogram: predicted total GC [Mg/Fe] distribution by
mass for the $<1 R_{\rm eff}$ shell, for a case in which a second episode of
star formation, induced by a gaseous merger, is taken into account (see
text). Solid histogram: observations by \cite{puzia06}, their entire
sample.}

\label{GD_mgfe2}

\end{figure}

We show our results in Figure~\ref{GD_merg} and \ref{GD_mgfe2}. We notice
a clear change in the overall shape of the metallicity distribution
$\Upsilon_{\rm GC, tot}$ with respect to the cases shown in the previous
sections, in the sense that now $\Upsilon_{\rm GC, tot}$ is narrower and
dominated by objects with super-solar metallicity (and sub-solar [Mg/Fe]
ratios) with a dominant population at [Z/H]~$\approx0.1$,  which is not
prominent in the observations of P06. The high-metallicity globular
cluster populations dominate the metallicity distribution which is at
variance with both the results from previous sections and the
observations.

{Our merger model does not include the accretion of globular clusters
that were already formed within the accreted satellite galaxies. The
inclusion of this effect could remedy the match between models and
observations at low metallicities, as the typical GC in a dwarf galaxy is
metal-poor \citep[e.g.][]{lotz04, sharina05} and their addition to the
initial GC population would enhance the total number of metal-poor GCs and
improve the fit to the data. However, these clusters need to be
$\alpha$-enhanced to match the observations. The impact of GC accretion on
our post-merger model predictions will be studied in detail in a future
paper. Here, we remark that the time at which the purely gaseous
subsequent merger event can occur (which does not import already formed
GCs), is limited by the onset of the galactic wind, after which the type-Ia SNe dominates the nucleosynthesis,
and needs to be completed at
$t_{\rm mrg} \la \!1-2$ Gyr after
the first starburst. However, this time constraint implies that such
merger events would overlap with the initial starburst and be mostly
indistinguishable from each other. Such a scenario closely resembles the
Searle-Zinn scenario \citep{sz78}, in which galaxy halos are formed from
the agglomeration of gaseous protogalactic fragments. Later merger 
events are excluded in our models, as they would produce GCs with 
sub-solar [Mg/Fe] ratios which is at variance with the observations.}

Note also that the fraction of GCs at [Z/H]$< -1$ can be recovered in our
models only if the cluster formation at low metallicity is enhanced (e.g.
using a value of 10 instead of 2 in eq.\ref{metaldepnd}). However, even in
the case in which we adopt some $f(Z)$ strongly declining with total
metallicity, which may alter the shape of $\Upsilon_{\rm GC, tot}$
enhancing the low-metallicity tail and thus improving the agreement with
observations, the position of both super-solar metallicity peaks will not
change, remaining at variance with the data.

\subsection{Other Mechanisms responsible for Multimodality}

\label{othermechs}

The picture emerging from our analysis is far from being the general
solution to explaining the complexity of GC color distributions, and it
suggests only a scheme in which multiple mechanisms could be at work
together, either broadening or adding features to the observed
distributions. 

\begin{figure}



\plotone{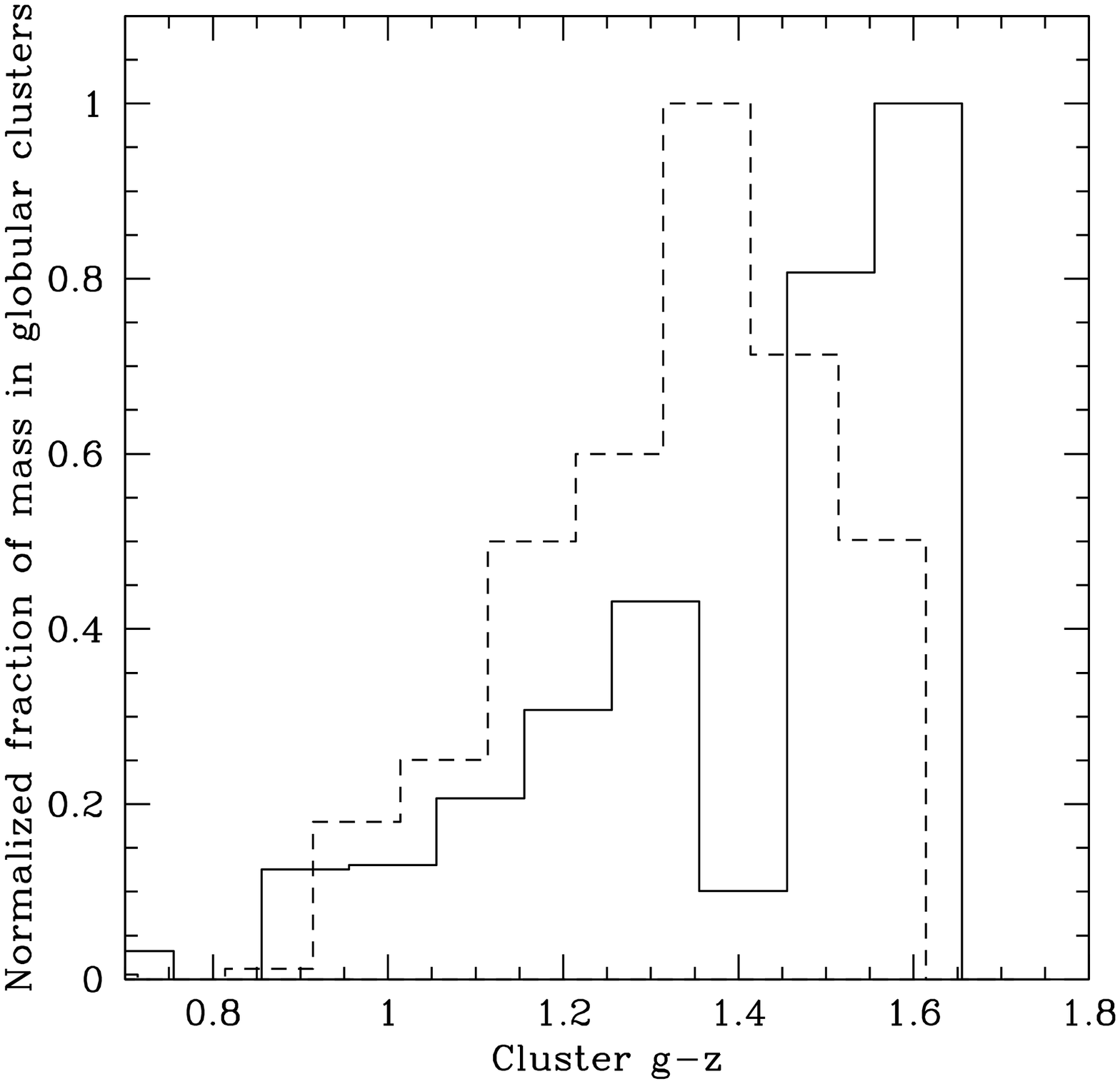}

\caption{Predicted globular-cluster metallicity distribution by mass as a function
of the $(g\!-\!z)$ colour at two different projected galactocentric radii.
Dashed: galactic core. Solid: galactic halo out to 10 $R_{\rm eff}$.} 
\label{yoon}

\end{figure}

For instance, \cite{yoon06} suggested that the color
bimodality could arise from the presence of hot horizontal-branch stars
(so far not accounted for in SSP model predictions) that results in a
non-linear color-metallicity transformation producing two color peaks from
an originally single-peak metallicity distribution. We tested this
scenario on our fiducial model { GC metallicity distribution}, by
applying to each SSP the following transformation from [Fe/H] to the
$(g-z)$ color:
\begin{eqnarray}
(g-z) = \alpha	+ \beta\,{\rm [Fe/H]} +\gamma \,{\rm [Fe/H]}^2 \\\nonumber
				+\delta \,{\rm [Fe/H}]^3 +\epsilon \,{\rm [Fe/H]}^4
\label{eq_yoon}
\end{eqnarray}
The numerical values of the coefficients are given in
Table~\ref{tab:coeff} and the relation was adopted from \cite{yoon06} and
is consistent with the best-fit relation presented in their Figure~1b. We
show our results in Figure~\ref{yoon}.

\begin{deluxetable}{cr}
\tabletypesize{\scriptsize}

\tablecaption{Numerical values of coefficients used in 
Equation~\ref{eq_yoon}. \label{tab:coeff}}

\tablewidth{0pt}

\tablehead{
\colhead{coefficient} & \colhead{numerical value}
}

\startdata

$\alpha$  	&	1.5033\\
$\beta$		&	0.172774\\
$\gamma$	& 	$-0.623522$\\
$\delta$		&	$-0.453331$\\
$\epsilon$	&	$-0.089038$\\
\enddata
\end{deluxetable}

Since we start from symmetric metallicity distributions, the non-linear
transformation seems to work and produce a color bimodality for the CSP
inhabiting the $< 10 R_{\rm eff}$ shell (Figure \ref{yoon}, solid line),
although the bimodality is slightly exaggerated compared to real data (see
Figure~\ref{colsamp}). In fact, a look at the color distribution which we
obtained for the sole $0.1 R_{\rm eff}$ shell reveals that it still has
one peak and is roughly symmetric (Figure \ref{yoon}, dashed line).
Obviously, since the $(g-z)-\!$~[Fe/H] relationship is meant to explain the
GCs color bimodality without invoking any other effect, we did not combine
the two histograms, either according to Eq.\ref{gcmd1} or to Eq.
\ref{gcmd2} in our models, as we are comparing metallicity distributions
to spectroscopic measurements. 

It is of great importance to investigate this transformation with large
and homogeneous data sets that cover a wide enough metallicity range to
allow a robust analysis of the non-linear inflection point in the
color-metallicity transformation. However, { as a result of
Figure~\ref{yoon}}, we point out that the color bimodality typically found
for GCSs in massive early-type galaxies might be { only} partly due to
a non-linear color-metallicity transformation.

{ Another effect put forward by, e.g.,\cite{recchi05} is the claim} that GCs
might have undergone a self-enrichment phase at the early stages of their
formation, and therefore some GCs could have experienced a boost in
metallicity which would be not representative of the metallicity of their
parent gas cloud. Finally, { as already mentioned above in
Section~\ref{ln:merger},} some GCs residing in the outermost regions of
the galaxies \citep[e.g.][]{lee06} could have experienced entirely
different chemical enrichment histories at the time of their formation and
later been added to a more massive system through accretion
\citep[e.g.][]{cote98}. The inclusion of these effects goes far beyond the
scope of this work, but we remind the reader that all the aforementioned
effects might influence the interpretation of any globular cluster color
and metallicity distribution.


\section{Conclusions}
By means of the comparison between PM04's best model predictions for the
radial changes in the CSP chemical properties and the recent spectroscopic
data on the metallicity distributions of extragalactic GCSs from
\cite{puzia06}, we are able to derive some conclusions on the GC
metallicity distributions in massive elliptical galaxies. In particular,
we focused on the main drivers of the { multi-modality} that is
observed in the majority of GCSs in massive elliptical galaxies. Our main
conclusions are:

\begin{itemize}

\item We show that the observed multi-modality in the GC
metallicity distributions can be ascribed to the radial variation in the
underlying stellar populations in giant elliptical galaxies. In
particular, the observed GCSs are consistent with a
linear combination of the GC sub-populations inhabiting
different galactocentric radii projected on the sky. 

\item A new prediction of our models, which is in astonishing agreement
with the spectroscopic observations, is the presence of a super-solar
metallicity mode that seems to emerge in the most massive elliptical
galaxies. In smaller objects, instead, this mode disappears quickly with
decreasing stellar mass of the host galaxy.

\item { Our models successfully reproduce the observed [Mg/Fe] bimodality
in GCSs of massive elliptical galaxies. This, in turn, suggests a
bimodality in formation timescales during the early formation epochs of GCs in
massive galaxy halos. The two modes are consistent with an early (initial) and later
(triggered) formation mode.}

\item {Since the GC populations trace the properties of galactic CSPs in our scenario,}
we predict an increase of the mean metallicity of the cluster system with
the host galaxy mass, which closely follows the mass-metallicity relation
for ellipticals. Moreover, we expect that a major fraction of the GCs
(i.e.~those born inside the galaxy) follows an age-metalliticity
relationship, in the sense that the older ones are also more
$\alpha$-enhanced and more metal-poor.

\item The role of host galaxy metallicity in shaping the observed GC 
metallicity distribution is non-negligible, although its effects
have been estimated to change the function $f\simeq \psi_{GC}/\psi_*$ by a
factor of $\sim 2-5$, in order to match the sample of \cite{puzia06}.
Either a non-linear color-metallicity transformation, or a stronger
metallicity effect, and/or accretion of GCs from the surrounding
environment is needed to explain a ratio of metal-poor to
metal-rich GCs close to unity, as reported for ellipticals based on
results from photometric surveys.

\item Merger models which include the later accretion of primordial and/or
solar-metallicity gas predict a shape for the GC metallicity
distribution which is at variance with the spectroscopic observations.

\end{itemize}

\acknowledgments

We thank the referee for a careful reading of the paper.
A.P. warmly thanks S.Recchi for useful discussions. A.P. acknowledges
support by the Italian Ministry for University under the COFIN03 prot.
2003028039 scheme. T.H.P. acknowledges support by NASA through grants
GO-10129 and GO-10515 from the Space Telescope Science Institute, which is
operated by AURA, Inc.,~under NASA Contract NAS5-26555, and the support in
form of a Plaskett Research Fellowship at the Herzberg Institute of
Astrophysics.

\clearpage

\end{document}